\documentclass{PoS}

\title{Non-extensive statistics effects in transverse momentum spectra of hadrons}

\ShortTitle{Non-extensive statistics effects in transverse momentum spectra of hadrons}

\author{\speaker{A.S.~Parvan} \\
        Bogoliubov Laboratory of Theoretical Physics, Joint Institute for Nuclear Research, 141980 Dubna, Russian Federation \\
        Department of Theoretical Physics, National Institute of Physics and Nuclear Engineering, P.O. Box MG-6, 077125 Bucharest-Magurele, Romania \\
        Institute of Applied Physics, Moldova Academy of Sciences, MD-2028 Chisinau, Republic of Moldova \\
        E-mail: \email{parvan@theor.jinr.ru}}

\abstract{The Bose-Einstein and Fermi-Dirac statistics of the identified hadrons were verified on the basis of the transverse momentum distributions of bosons and fermions created in the $pp$ collisions at high energies using the Tsallis-factorized   statistics and the two-component distributions of the Boltzmann-Gibbs statistics. The main differences between the parameters of the Tsallis-factorized statistics and the Boltzmann-Gibbs statistics were identified. The results of the Boltzmann-Gibbs statistics are reasonable and suggest that the soft and hard hadrons may be produced from the two different macroscopic states of the dynamical system. It was revealed that the volume of the system obtained in the Tsallis-factorized statistics is unusually large in comparison with the geometrical volume of two protons. The main formulas for the Tsallis statistics in the grand canonical ensemble were formulated.}

\FullConference{XXII International Baldin Seminar on High Energy Physics Problems \\
                 15-20 September, 2014\\
                 JINR, Dubna, Russia}

\begin{document}

\section{Introduction}~\label{sec1}
The measurement of the transverse momentum and rapidity distributions of final-state particles produced in the hadron-hadron and nucleus-nucleus collisions is considered to provide important information about the collision process and the particle production mechanism~\cite{Florkowski}. The transverse momentum distributions of identified hadrons are the most common tools used to investigate strong interactions leading to the creation of particles in the high-energy collisions due to the momentum balance in the plane perpendicular to the beam direction. The missing transverse energy is a key quantity in many searches for physics beyond the Standard Model, such as supersymmetry, extra dimensions and dark matter, and in the study of the Higgs boson, neutrinos, $W$ bosons and top quarks~\cite{Cms2}.

In the modern paradigm of science all known elementary particles are governed by the Bose-Einstein or Fermi-Dirac statistics, i.e., they are bosons or fermions. The noninteracting bosons and fermions given in the framework of the Boltzmann-Gibbs statistics under the constraints of the grand canonical ensemble are described by the Bose-Einstein and Fermi-Dirac distributions, $f=1/[\exp((\varepsilon-\mu)/T)\pm 1]$. These laws are followed by the photons irradiated in atomic and nuclear phenomena and by the particles emitted in the hadron-hadron and heavy-ion collisions at low and intermediate energies~\cite{Yagi}. However, the transverse momentum distributions of the particles produced in the $pp$ and $AA$ collisions at LHC and RHIC energies can not be described by the Bose-Einstein and the Fermi-Dirac distributions of the Boltzmann-Gibbs statistics given in the grand canonical ensemble. Thus, if we continue to accept that the particles at high-energies are bosons or fermions described by the Bose-Einstein or the Fermi-Dirac statistics, respectively, then we should find causes of changes of their distribution functions.

The main purpose of this paper is to verify the statistics of identified hadrons produced in $pp$ collisions at high energies and to identify possible scenarios for the creation of hard hadrons with a large transverse momentum. Thereby, we investigate the data using the Bose-Einstein and Fermi-Dirac distributions of the Tsallis-factorized   statistics and introducing the two-component Bose-Einstein and Fermi-Dirac distributions of the ordinary Boltzmann-Gibbs statistics.

In the analysis of the LHC and RHIC data, a Tsallis-like distribution gives satisfactory fits to the transverse momentum distributions~\cite{Alice1,Alice1a,Alice1b,Atlas1,Cms3}. However, hitherto in most works the transverse momentum distributions were fitted only by the classical Maxwell-Boltzmann distribution in both the Tsallis-factorized statistics~\cite{Rybczynski14,Cleymans13,Azmi14,Cleymans12a,Cleymans12b,Marques13} and the two sources Boltzmann-Gibbs statistics~\cite{Liu14,Li14}.

The structure of the paper is as follows. In Section~\ref{sec2}, we briefly define the main formulas of the Tsallis statistics in the grand canonical ensemble. In Section~\ref{sec3}, we remind the transverse momentum and rapidity distributions in the Tsallis-factorized statistics. The analysis of the data and the results are discussed in Section~\ref{sec4}. The main conclusions are summarized in the final section.

\section{Tsallis statistics in grand canonical ensemble}~\label{sec2}
Let us consider the equilibrium statistical ensemble of the dynamical systems at the constant temperature $T$, volume $V$, chemical potential $\mu$ and the thermodynamic coordinate $z$ in a thermal contact with a heat bath. The system exchanges the energy and the matter with its surroundings. In order to determine the probabilities of microstates, we consider the Tsallis statistical entropy~\cite{Tsallis88} which is a function of the parameter $q$ and a functional of the probing probabilities $p_{i}$:
\begin{equation}\label{1}
    S=-k_{B} \sum\limits_{i} \frac{p_{i}-p_{i}^{q}}{1-q}=k_{B} z \sum\limits_{i} p_{i} (1-p_{i}^{1/z}),
\end{equation}
where $k_{B}$ is the Boltzmann constant, $z=1/(q-1)$ and $q\in\mathbf{R}$ is a real parameter taking values $0<q<\infty$. In the limit $q\to 1$ the entropy (\ref{1}) recovers the Boltzmann-Gibbs entropy, $S_{G}=-k_{B} \sum\limits_{i} p_{i} \ln p_{i}$. The probabilities of microstates are normalized to unity:
\begin{equation}\label{2}
    \sum\limits_{i} p_{i} =1.
\end{equation}
The parameter $z$ takes the values $-\infty < z < -1$ for $0<q < 1$ and $0<z<\infty$ for $1< q<\infty$. The expectation values of the Hamiltonian and the number of particles can be written as
\begin{eqnarray}\label{3}
    \langle H\rangle &=& \sum\limits_{i} p_{i} E_{i}, \\ \label{4}
     \langle N\rangle &=& \sum\limits_{i} p_{i} N_{i},
\end{eqnarray}
where $E_{i}$ and $N_{i}$ are the energy and number of particles in the $i$-th microstate of the system. In the state of thermal equilibrium the probabilities of microstates of the grand canonical ensemble can be found from the maximum entropy principle by the fundamental equation of thermodynamics at the fixed values of the variables of state $(T,V,z,\mu)$ or by the Jaynes principle:
\begin{eqnarray}\label{8}
p_{i} &=& \left[1+\frac{1}{z+1}\frac{\Lambda-E_{i}+\mu N_{i}}{k_{B}T}\right]^{z}, \\ \label{9}
    1 &=& \sum\limits_{i} \left[1+\frac{1}{z+1}\frac{\Lambda-E_{i}+\mu N_{i}}{k_{B}T}\right]^{z},
\end{eqnarray}
where $\Lambda$ is determined from Eq.~(\ref{9}) and it is a function of the state variables, $\Lambda=\Lambda(T,V,z,N)$. In the limit $q\to 1$ $(z \to\pm\infty)$ the probability $p_{i}^{(G)}=\exp[(\Omega_{G}-E_{i}+\mu N_{i})/k_{B}T]$ and the function $\Lambda$ recovers the grand potential $\Omega_{G}=-k_{B}T\ln Z_{G}$, where $Z_{G}=\sum_{i} \exp[-(E_{i}-\mu N_{i})/k_{B}T]$. The expectation value $\langle A \rangle$ of the operator $A$ can be defined as follows:
\begin{equation}\label{10}
\langle A \rangle = \sum\limits_{i} A_{i} \left[1+\frac{1}{z+1}\frac{\Lambda-E_{i}+\mu N_{i}}{k_{B}T}\right]^{z}.
\end{equation}

In the Tsallis statistics, the mean occupation numbers for the Bose-Einsten, Fermi-Dirac and Maxwell-Boltzmann statistics of noninteracting particles in the grand canonical ensemble can be defined as
\begin{equation}\label{11}
 \langle n_{\vec{p}\sigma} \rangle = \sum\limits_{\{n_{\vec{p}\sigma}\}} G\{n_{\vec{p}\sigma}\} n_{\vec{p}\sigma} \left[1+\frac{1}{z+1} \frac{\Lambda-\sum_{\vec{p}\sigma}n_{\vec{p}\sigma}(\varepsilon_{\vec{p}}-\mu)}{k_{B}T} \right]^{z},
\end{equation}
where $\varepsilon_{\vec{p}}=\sqrt{\vec{p}^{2}+m^{2}}$ is the particle energy, $G\{n_{\vec{p}\sigma}\}=1$ for the Fermi-Dirac statistics ($n_{\vec{p}\sigma}=0,1$) and the Bose-Einstein statistics ($n_{\vec{p}\sigma}=0,1,\ldots,\infty$) of particles and $G\{n_{\vec{p}\sigma}\}=1/\prod_{\vec{p}\sigma}n_{\vec{p}\sigma}!$ for the Maxwell-Boltzmann statistics ($n_{\vec{p}\sigma}=0,1,\ldots,\infty$) of particles~\cite{Parvan04}.

The transverse momentum and rapidity distributions of particles can be defined as
\begin{equation}\label{17}
  \frac{d^{2}N}{dp_{T}dy} = \frac{V}{h^{3}}  \int\limits_{0}^{2\pi} d\varphi p_{T} \varepsilon_{\vec{p}} \sum\limits_{\sigma} \langle n_{\vec{p}\sigma}\rangle
\end{equation}
and
\begin{equation}\label{18}
  \frac{dN}{dy} = \frac{V}{h^{3}} \int\limits_{0}^{2\pi} d\varphi \int\limits_{0}^{\infty} dp_{T} p_{T} \varepsilon_{\vec{p}} \sum\limits_{\sigma} \langle n_{\vec{p}\sigma}\rangle,
\end{equation}
where $p_{T}$ and $y$ are the transverse momentum and rapidity, respectively, and $\langle n_{\vec{p}\sigma}\rangle$ are the mean occupation numbers given by Eq.~(\ref{11}).

\section{Tsallis-factorized   statistics}~\label{sec3}
The Tsallis-factorized   statistics is defined by the generalized entropy for the ideal gas in the grand canonical ensemble~\cite{Cleymans12b}. In the Boltzmann-Gibbs statistics the entropy of the ideal gas in the grand canonical ensemble can be written as
\begin{equation}\label{19}
  S_{G}=-g \sum\limits_{\vec{p}} \left[ \langle n_{\vec{p}}\rangle_{G} \ln \langle n_{\vec{p}}\rangle_{G} +\frac{1}{\eta} (1-\eta \langle n_{\vec{p}}\rangle_{G}) \ln (1-\eta \langle n_{\vec{p}}\rangle_{G})\right],
\end{equation}
where $\langle n_{\vec{p}}\rangle_{G}=1/(\exp[\beta (\varepsilon_{\vec{p}}-\mu)]+\eta)$, $\eta=-1$ for the Bose-Einstein statistics, $\eta=0$ for the Maxwell-Boltzmann statistics and $\eta=1$ for the Fermi-Dirac statistics of particles. In the Tsallis-factorized   statistics, the Boltzmann-Gibbs entropy (\ref{19}) for the Bose-Einstein and the Fermi-Dirac statistics is rewritten, using the Tsallis prescription, in the following form~\cite{Buyukkilic93,Pennini95,Teweldeberhan04}
\begin{equation}\label{20}
  S=-g \sum\limits_{\vec{p}} \left[ \langle n_{\vec{p}}\rangle^{q} \ln_{q} \langle n_{\vec{p}}\rangle + \frac{1}{\eta} (1-\eta \langle n_{\vec{p}}\rangle)^{q} \ln_{q} (1-\eta \langle n_{\vec{p}}\rangle)\right],
\end{equation}
where
\begin{equation}\label{21}
  \ln_{q}(x) = \frac{x^{1-q}-1}{1-q}
\end{equation}
and $q$ is a real parameter, $q>0$. For the Maxwell-Boltzmann statistics, the Boltzmann-Gibbs entropy (\ref{19}) is rewritten as~\cite{Biro11}
\begin{equation}\label{22}
   S=-g \sum\limits_{\vec{p}} \left[ \langle n_{\vec{p}}\rangle^{q} \ln_{q} \langle n_{\vec{p}}\rangle - \langle n_{\vec{p}}\rangle\right].
\end{equation}
The expectation value of the Hamiltonian and the mean number of particles of the system in the Tsallis-factorized   statistics are defined as~\cite{Cleymans12b}
\begin{eqnarray}\label{23}
    \langle H\rangle &=& g\sum\limits_{\vec{p}} \varepsilon_{\vec{p}} \langle n_{\vec{p}}\rangle^{q}, \\ \label{24}
     \langle N\rangle &=& g\sum\limits_{\vec{p}} \langle n_{\vec{p}}\rangle^{q},
\end{eqnarray}
where $E_{i}$ and $N_{i}$ are the energy and the number of particles in the $i$-th microstate of the system. Maximization of the entropies (\ref{20}) and (\ref{22}) under the constraints (\ref{23}) and (\ref{24}) leads to the mean occupation numbers for the Tsallis-factorized   statistics~\cite{Cleymans12b}
\begin{equation}\label{25}
  \langle n_{\vec{p}}\rangle = \frac{1}{\left[1+(q-1) \beta (\varepsilon_{\vec{p}}-\mu) \right]^{\frac{1}{q-1}}+\eta},
\end{equation}
where $\eta=-1$ for the Bose-Einstein statistics, $\eta=0$ for the Maxwell-Boltzmann statistics and $\eta=1$ for the Fermi-Dirac statistics of particles. Then the transverse momentum and rapidity distributions of particles can be written as~\cite{Cleymans12b}
\begin{eqnarray}\label{26}
  \frac{d^{2}N}{dp_{T}dy} &=& \frac{gV}{h^{3}}  \int\limits_{0}^{2\pi} d\varphi p_{T} \varepsilon_{\vec{p}} \langle n_{\vec{p}}\rangle^{q} \nonumber \\
  &=& \frac{gV}{(2\pi)^{2}} \frac{ p_{T}  m_{T} \cosh y}{\left(\left[1+(q-1) \beta ( m_{T} \cosh y-\mu) \right]^{\frac{1}{q-1}}+\eta\right)^{q}}
\end{eqnarray}
and
\begin{eqnarray}\label{27}
  \frac{dN}{dy} &=& \frac{gV}{h^{3}} \int\limits_{0}^{2\pi} d\varphi \int\limits_{0}^{\infty} dp_{T} p_{T} \varepsilon_{\vec{p}} \langle n_{\vec{p}}\rangle^{q} \nonumber \\
  &=& \frac{gV}{(2\pi)^{2}} \int\limits_{0}^{\infty}  p_{T}  dp_{T} \frac{m_{T} \cosh y}{\left(\left[1+(q-1) \beta ( m_{T} \cosh y-\mu) \right]^{\frac{1}{q-1}}+\eta\right)^{q}},
\end{eqnarray}
where $m_{T}=\sqrt{p_{T}^{2}+m^{2}}$. Note that throughout the paper we use the system of natural units, $\hbar=c=k_{B}=1$.

\section{Analysis of transverse momentum distributions}~\label{sec4}
Let us verify that bosons and fermions produced in the $pp$ collisions at LHC energies are described by the corresponding Bose-Einstein and Fermi-Dirac functions of both the Boltzmann-Gibbs statistics and the Tsallis-factorized   statistics. We should check the fact that the particles created in the violent high-energy collisions are indeed bosons or fermions as those produced at lower energies.

The transverse momentum distributions for the bosons ($\eta=-1$), fermions ($\eta=1$) and Boltzmann particles ($\eta=0$) in the Boltzmann-Gibbs statistics are given by
\begin{equation}\label{28}
  \frac{d^{2}N}{dp_{T}dy} = \frac{gV}{(2\pi)^{2}}  p_{T}  m_{T} \cosh y \ \frac{1}{e^{\frac{ m_{T} \cosh y-\mu}{T}}  +\eta}.
\end{equation}
These distributions correspond to the relativistic Bose-Einstein, Fermi-Dirac or Maxwell-Boltzmann ideal gases composed of many randomly moving identical point particles in equilibrium, which do not interact and which are given in the grand canonical ensemble at temperature $T$ and chemical potential $\mu$ in the volume $V$ with imaginary walls. The chemical potential of antiparticles in Eq.~(\ref{28}) is taken with the opposite sign. At high-energies of the LHC the baryon chemical potential is zero, i.e. $\mu=0$.

The creation of particles (bosons and fermions) in the high-energy $pp$ collisions is a dynamical process. However, the transverse momentum distributions (\ref{17}), (\ref{26}) and (\ref{28}) of the Tsallis, Tsallis-factorized and the Boltzmann-Gibbs statistics can be applied only for the equilibrium systems because they were obtained from the maximum entropy principle. Thus, the description of particle production process in high-energy collisions by these distributions leads to the substitution of the complex dynamical system with an equilibrium system of ideal gases. Is it possible to describe such a complex dynamic process of creation of many particles only with a one equilibrium state of the system? Do all the particles produced in high-energy collisions originate from the same macroscopic state of the system or from different macroscopic states corresponding to different intervals of time of collision?

Let us introduce the two-component transverse momentum distributions for the bosons, fermions and boltzmann particles given in the framework of the Boltzmann-Gibbs statistics
\begin{equation}\label{29}
  \frac{d^{2}N}{dp_{T}dy} = \frac{gV_{1}}{(2\pi)^{2}}  p_{T} m_{T}\cosh y \ \frac{1}{e^{\frac{ m_{T} \cosh y-\mu_{1}}{T_{1}}}  +\eta} +
  \frac{gV_{2}}{(2\pi)^{2}} p_{T} m_{T}\cosh y \ \frac{1}{e^{\frac{ m_{T} \cosh y-\mu_{2}}{T_{2}}}  +\eta}.
\end{equation}
This distribution corresponds to the ideal gas in two different macrostates of the equilibrium system (two sources). The first term describes the particles created in the macrostate $(T_{1},V_{1},\mu_{1})$ and the second term is related to the production of the same type of particles from the other macrostate $(T_{2},V_{2},\mu_{2})$. The parameters $(T,V,\mu)$ evolve during the evolution of the system.

\begin{figure}
\includegraphics[width=14cm]{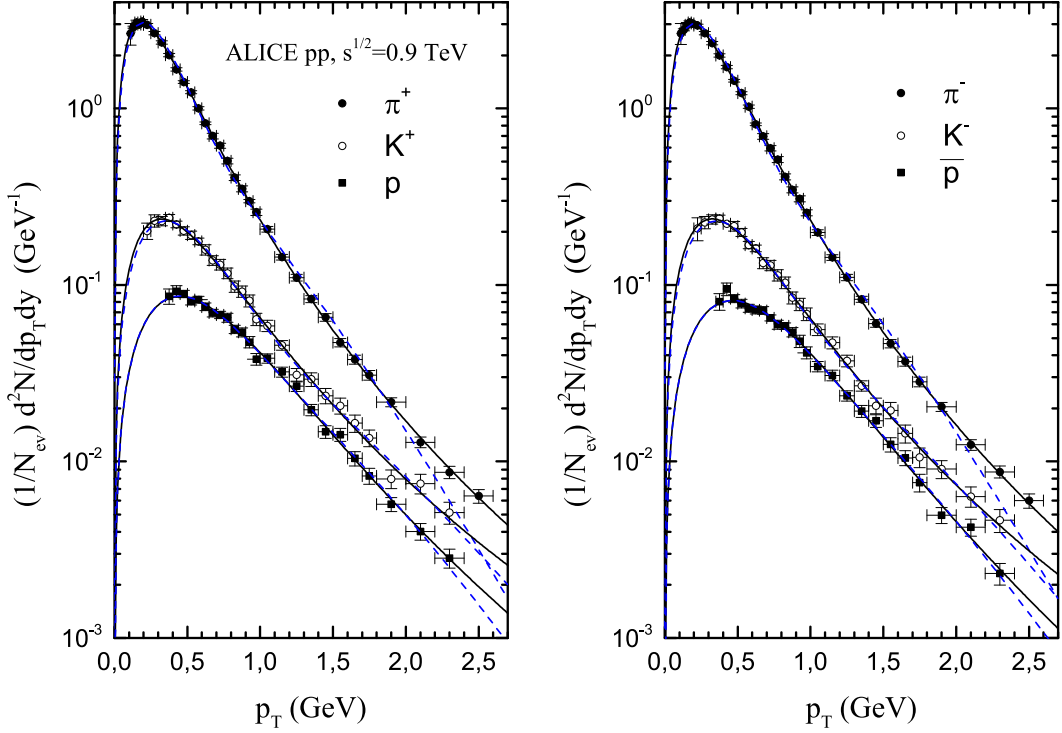} \vspace{-0.3cm}
\caption{(Color online) Transverse momentum distributions of identified hadrons produced in the $pp$ collisions as obtained by the ALICE Collaboration~\cite{Alice1} at $\sqrt{s}=0.9$ TeV. The solid and dashed curves are the fits of the data to the Tsallis-factorized distributions and the two-component Boltzmann-Gibbs distributions, respectively, at rapidity $y=0$. }   \label{fig1}
\end{figure}
Let us analyze the transverse momentum spectra of particles created in the $pp$ collisions at LHC energies. Figure~1 represents the transverse momentum distributions of pions $\pi^{\pm}$, kaons $K^{\pm}$, protons and antiprotons produced in the $pp$ collisions at energy $\sqrt{s}=0.9$ TeV. The symbols represent the experimental data of the ALICE Collaboration~\cite{Alice1}. The solid and the dashed curves are the fits of the experimental data to the Tsallis-factorized function (\ref{26}) and the two-component Boltzmann-Gibbs function (\ref{29}), respectively, at rapidity $y=0$. The values of the variables of state for the macrostates of the system for the Tsallis-factorized statistics and for the Boltzmann-Gibbs statistics are given in Tables~1 and 2, respectively. The pions and kaons follow the Bose-Einstein statistics. The protons and antiprotons follow the Fermi-Dirac statistics. The Bose-Einstein and Fermi-Dirac transverse momentum functions of the Tsallis-factorized   statistics describe the experimental data very well. The values of the Tsallis-factorized temperature $T$, radius $R$ of the system and the parameter $q$ fluctuate in the dependence on the type of a hadron. The proton temperature is much lower than the temperature of other hadrons; however, the volume of the system for protons is much larger than the volume of the system for other hadrons. Nevertheless, the value of the parameter $q$ for protons is consistent with the values of the parameter $q$ of other types of hadrons. In the $pp$ collisions, the Tsallis-factorized volume of the system for all types of hadrons considerably overestimates the realistic volume of the system of two protons. The two-component Bose-Einstein and Fermi-Dirac transverse momentum functions of the Boltzmann-Gibbs statistics describe very well only kaons, protons and antiprotons. But the high-$p_{T}$ pions failed to be described by these two-component distributions. For all types of hadrons the two-component distribution is defined by two different sets of parameters $(T,R)$ corresponding to two macroscopic states of the dynamic system. The first set of parameters corresponding to the lower part of $p_{T}$-spectra (soft) is characterized by smaller values of the temperature and larger values of the volume in comparison with the same parameters from the second set. The second set of parameters corresponding to the higher part of the $p_{T}$-spectra (hard) is characterized by high temperatures and small volumes. The radius $R_{1}$ is consistent with the radius of the system of two protons;  however, the radius $R_{2}$ corresponds to a strongly compressed system of two protons (core). Thus, in the $pp$ collisions the high-$p_{T}$ hadrons are created in a system of a very small volume and large temperature. However, the low-$p_{T}$ hadrons are produced in a system of larger volume and smaller temperature. The Gibbs temperature $T_{1}$ is about twice the Tsallis-factorized   temperature $T$ and the temperature $T_{2}$ is about twice the temperature $T_{1}$.

\begin{figure}
\includegraphics[width=8cm]{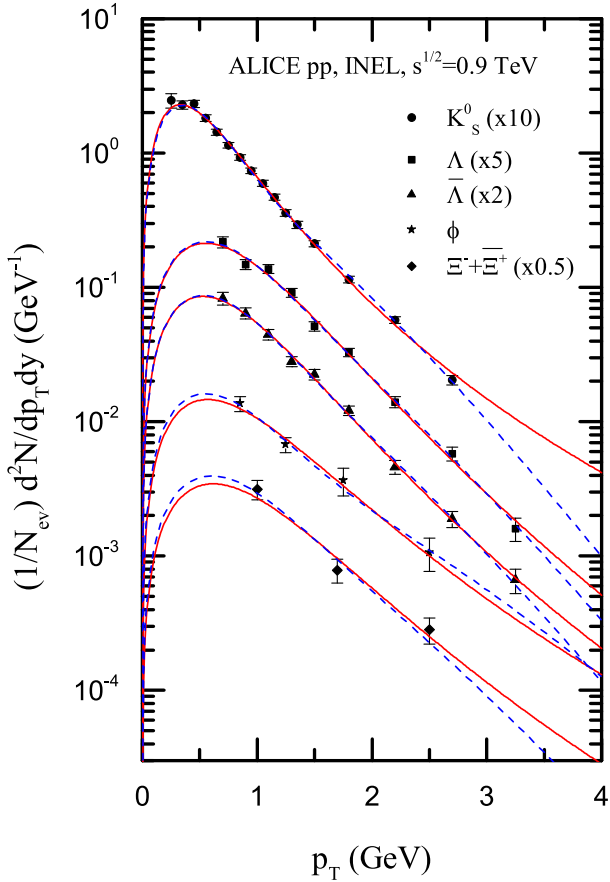} \vspace{-0.3cm}
\caption{(Color online) Transverse momentum distributions of identified hadrons produced in the $pp$ collisions as obtained by the ALICE Collaboration~\cite{Alice2} at $\sqrt{s}=0.9$ TeV. The solid and dashed curves are the fits of the data to the Tsallis-factorized   distributions and the two-component Boltzmann-Gibbs distributions, respectively, at rapidity $y=0$.} \label{fig2}
\end{figure}
Figure~2 represents the transverse momentum distributions of the identified hadrons produced in the $pp$ collisions at energy $\sqrt{s}=0.9$ TeV. The symbols represent the experimental data of the ALICE Collaboration~\cite{Alice2}. The solid curves are the fits of the experimental data to the Tsallis-factorized   distributions (\ref{26}) and the dashed curves are the fits of the experimental data to the two-component Bose-Einstein and Fermi-Dirac transverse momentum functions of the Boltzmann-Gibbs statistics(\ref{29}) at rapidity $y=0$. The values of the variables of state for the Tsallis-factorized   statistics and for the Boltzmann-Gibbs statistics are given in Tables~1 and 2, respectively. The mesons $K_{S}^{0}$ and $\phi$ follow the Bose-Einstein statistics and the strange baryons follow the Fermi-Dirac statistics. The Bose-Einstein and Fermi-Dirac transverse momentum distributions of the Tsallis-factorized   statistics describe the experimental data very well. The Tsallis-factorized temperature $T$ and radius $R$ of the system strongly fluctuate in the dependence on the hadron type. The value of the parameter $q$ fluctuates more slightly. The temperature of the strange baryon $\Lambda$ differs essentially from the temperature of its antiparticle. The Tsallis-factorized volume of the system for $K_{S}^{0}$, $\overline{\Lambda}$ and $\Xi^{-}+\overline{\Xi}^{+}$ considerably disagrees with the realistic volume of the system of two protons. The two-component Bose-Einstein and Fermi-Dirac transverse momentum distributions of the Boltzmann-Gibbs statistics describe the experimental data for all these identified hadrons very well. For the Boltzmann-Gibbs distributions the values of the temperature and the radius of the system also fluctuate considerably in the dependence on the type of a hadron. The first temperature $T_{1}$ is smaller than the second temperature $T_{2}$ and the first radius $R_{1}$ is larger than the second radius $R_{2}$. The first set of parameters given by the small value of temperature and the large value of volume describes the low values of the transverse momentum spectra; however, the second set of parameters given by the large value of temperature and the very small value of volume describes the high values of the transverse momentum spectra. The radius $R_{1}$ is consistent with the radius of the system of two protons and the radius $R_{2}$ corresponds to a compressed system of two protons. From the two-component Boltzmann-Gibbs distributions it is seen that the high-$p_{T}$ hadrons are produced from  a very compressed and heated system. The Gibbs temperature $T_{1}$ is larger than the Tsallis-factorized   temperature $T$ and the Gibbs temperature $T_{2}$ is larger than the temperature $T_{1}$.

\begin{table}
\begin{tabular}{cccccc}
 \hline
 \hline
 $\quad$ Type $\quad$         &$\quad T$, MeV$\qquad$ & $\qquad R$, fm$\qquad$ &$\qquad q\qquad$ & $\chi^{2}/\nu$ & Comments  \\
 \hline
 $\pi^{+}$                    & 75.94$\pm$0.57         & 4.397$\pm$0.032        & 1.142$\pm$0.001 & 0.49                       & Fig.~1   \\
 $\pi^{-}$                    & 78.34$\pm$0.52         & 4.262$\pm$0.027        & 1.137$\pm$0.001 & 0.39                       &          \\
 $K^{+}$                      & 58.32$\pm$6.22         & 4.413$\pm$0.608        & 1.172$\pm$0.007 & 0.55                       &          \\
 $K^{-}$                      & 62.75$\pm$4.27         & 4.070$\pm$0.348        & 1.164$\pm$0.005 & 0.30                       &          \\
 $p$                          & 33.44$\pm$13.2         & 16.365$\pm$12.103      & 1.147$\pm$0.009 & 0.66                       &           \\
 $\overline{p}$               & 62.68$\pm$13.3         & 5.449$\pm$1.991        & 1.123$\pm$0.009 & 0.65                       &           \\
 $K_{S}^{0}$                  & 69.86$\pm$6.38         & 3.531$\pm$0.441        & 1.158$\pm$0.006 & 0.78                       & Fig.~2    \\
 $\Lambda$                    & 104.79$\pm$34.00       & 2.367$\pm$1.379        & 1.096$\pm$0.017 & 1.59                       &            \\
 $\overline{\Lambda}$         & 74.99$\pm$24.5         & 4.273$\pm$2.688        & 1.111$\pm$0.012 & 0.62                       &            \\
 $\phi$                       & 96                     & 1.228$\pm$0.104        & 1.136$\pm$0.011 & 0.60                       &            \\
 $\Xi^{-}+\overline{\Xi}^{+}$ & 60                     & 3.221$\pm$0.176        & 1.13            & 2.21                       &            \\
\hline
\hline
\end{tabular}
\caption{Values of the parameters corresponding to the curves of fitting data to the Tsallis-factorized functions in Figs.~1-2. }
\label{t1}
\end{table}

In Tables~1 and 2 we have the values of the variables of state for the Tsallis-factorized statistics and the Boltzmann-Gibbs statistics, respectively, for each type of hadrons produced in the $pp$ collisions at $\sqrt{s}=0.9$ TeV measured by the ALICE Collaboration~\cite{Alice1,Alice2}. The values of the parameters are not the same for different types of hadrons. However, at the fixed energy of collision the values of the parameters should be independent of the hadron type. Thus, let us find the mean values of the parameters for the Tsallis-factorized statistics and the Boltzmann-Gibbs statistics and compare them. For the experimental data of the ALICE Collaboration~\cite{Alice1,Alice2}, the mean value of the Tsallis-factorized temperature $T=70.65\pm 4.27$ MeV, the mean value of the Tsallis-factorized radius of the system $R=4.871\pm 1.151$ fm and the mean value of the Tsallis-factorized parameter $q=1.138\pm 0.003$. For the two-component Bose-Einstein and Fermi-Dirac distributions of the Boltzmann-Gibbs statistics we have two sets of parameters. The mean value of the first temperature $T_{1}=148.38\pm 4.59$ MeV and the mean value of the first radius of the system $R_{1}=1.799\pm 0.091$ fm. The mean value of the second temperature $T_{2}=332.25\pm 12.15$ MeV and the mean value of the second radius of the system $R_{2}=0.419\pm 0.016$ fm.

\begin{table}
\begin{tabular}{cccccccc}
 \hline
 \hline
 Type &$\quad T_{1}$, MeV$\quad$& $\quad R_{1}$, fm$\quad$    &$\quad T_{2}$, MeV$\quad$ &$\quad R_{2}$, fm$\quad$ &$\chi^{2}/\nu$ & Comments\\
 \hline
 $\pi^{+}$                    & 102.26$\pm$1.96    & 3.424$\pm$0.059    & 250.28$\pm$6.26    & 0.934$\pm$0.044    & 5.79          & Fig.~1  \\
 $\pi^{-}$                    & 104.56$\pm$1.84    & 3.348$\pm$0.052    & 252.44$\pm$6.32    & 0.905$\pm$0.044    & 4.81          &         \\
 $K^{+}$                      & 142.30$\pm$4.93    & 1.671$\pm$0.077    & 345.81$\pm$17.4    & 0.380$\pm$0.035    & 0.60          &         \\
 $K^{-}$                      & 141.51$\pm$4.11    & 1.677$\pm$0.060    & 332.11$\pm$14.0    & 0.399$\pm$0.032    & 0.46	       &         \\
 $p$                          & 133.99$\pm$11.2    & 2.216$\pm$0.432    & 293.77$\pm$16.4    & 0.420$\pm$0.058    & 0.73          &         \\
 $\overline{p}$               & 142.60$\pm$13.3    & 1.860$\pm$0.366    & 291.06$\pm$20.0    & 0.416$\pm$0.074    & 0.71	       &         \\
 $K_{S}^{0}$                  & 141.12$\pm$5.10    & 1.687$\pm$0.099    & 340.85$\pm$8.77    & 0.393$\pm$0.021    & 0.75          & Fig.~2  \\
 $\Lambda$                    & 179.17$\pm$38.5    & 1.225$\pm$0.598    & 352.36$\pm$45.4    & 0.267$\pm$0.100    & 1.69          &         \\
 $\overline{\Lambda}$         & 170.38$\pm$26.3    & 1.385$\pm$0.523    & 348.92$\pm$32.0    & 0.265$\pm$0.070    & 0.81          &         \\
 $\phi$                       & 191.58	           & 0.597$\pm$0.061    & 473.51$\pm$116     & 0.104$\pm$0.045    & 0.66 	       &         \\
 $\Xi^{-}+\overline{\Xi}^{+}$ & 182.7              & 0.700$\pm$0.112    & 373.66             & 0.129$\pm$0.013    & 1.90	       &         \\
\hline
\hline
\end{tabular}
\caption{Values of the parameters corresponding to the curves of fitting data to the two-component Boltzmann-Gibbs functions in Figs.~1-2.}
\label{t2}
\end{table}

It is seen that in the $pp$ collisions at $\sqrt{s}=0.9$ TeV, the Boltzmann-Gibbs statistics suggests two regimes of production of particles. In the first regime, the Boltzmann-Gibbs soft hadrons are produced at the temperature $T_{1}\approx 150$ MeV from a heated spatial region of the volume consistent with the geometrical volume of two protons. However, in the second regime, the Boltzmann-Gibbs hard hadrons with a high transverse momentum originate from a hot superdense spatial region of the radius $R_{2}\approx 0.4$ fm with the temperature $T_{2}\approx 330$ MeV. In contrast, the Tsallis-factorized statistics presupposes that the particles in the $pp$ collisions at $\sqrt{s}=0.9$ TeV are created from a large nonphysical volume of the radius $R\approx 5$ fm at the small temperature $T\approx 70$ MeV.

\section{Discussion and conclusions}~\label{sec5}
In conclusion, the Bose-Einstein and Fermi-Dirac distributions of the Tsallis-factorized and the two-component Boltzmann-Gibbs statistics lead to a satisfactory description of the data on the transverse momentum for the identified hadrons created in the high energy $pp$ collisions. We have extracted the parameters $T,V$ and $q$ of the Tsallis-factorized   statistics and the two sets of parameters $T_{1},V_{1}$ and $T_{2},V_{2}$ of the two-component Boltzmann-Gibbs statistics for each type of hadrons measured by the ALICE Collaboration~\cite{Alice1,Alice2}. These parameters for different types of hadrons have different values for both the Tsallis-factorized statistics and the two-component Boltzmann-Gibbs statistics. The mean values of these parameters for the Tsallis-factorized   statistics and the two-component Boltzmann-Gibbs statistics were obtained. The results of the Boltzmann-Gibbs statistics for the $pp$ collisions at $\sqrt{s}=0.9$ TeV suggest that the soft hadrons with low transverse momentum are created from the spatial volume comparable with the geometrical volume of two protons at the temperature of $150$ MeV and the hard hadrons with high transverse momentum are produced from the very small volume in comparison with the volume of two protons at very high temperature of $330$ MeV. In contrast, the volume of the system obtained by the Tsallis-factorized   statistics is unusually large in comparison with the volume of two protons and the temperature is much less than both temperatures of the Boltzmann-Gibbs statistics.

Note that the high-$p_{T}$ tail of the charged pion distributions cannot be described by the two-component Boltzmann-Gibbs statistics. This discrepancy for pions in hadronic collisions can be understood as a superposition of different production sources.

{\bf Acknowledgments:} This work was supported in part by the joint research project of JINR and IFIN-HH, protocol N~4342.

\end{document}